\documentclass[11pt,twocolumn]{article}
\usepackage[margin=1in,columnsep=0.35in]{geometry}
\usepackage[utf8]{inputenc}
\usepackage[T1]{fontenc}
\usepackage{amsmath}
\usepackage{bm}         
\usepackage{graphicx}
\usepackage{url}
\usepackage{soul}
\usepackage{xcolor}
\usepackage{colortbl}   
\usepackage{array}
\usepackage{dcolumn}    
\usepackage{etoolbox}
\usepackage{comment}
\usepackage[version=4]{mhchem}
\usepackage{subcaption}
\usepackage{booktabs,tabularx,array} 
\usepackage{threeparttable} 
\usepackage[authoryear,round]{natbib}
\usepackage{authblk}
\usepackage[colorlinks=true,citecolor=blue,linkcolor=blue,urlcolor=blue]{hyperref} 

\usepackage{booktabs}
\usepackage{siunitx}
\usepackage{tabularx}

\usepackage[switch]{lineno}   
\makeatletter
\newsavebox{\abstractbox}
\renewenvironment{abstract}
  {\global\setbox\abstractbox\vbox\bgroup
   \hsize=0.9\textwidth \parindent=0pt \small
   \noindent\textbf{Abstract.}\ \ignorespaces}
  {\egroup}
\renewcommand\@maketitle{%
  \newpage\null\vskip 0.5em
  \begin{center}%
    {\large\bfseries \@title\par}%
    \vskip 0.8em
    {\@author\par}%
    \vskip 1em
    \usebox{\abstractbox}%
  \end{center}%
  \par\vskip 0.5em}
\makeatother
\let\cite\citep   
\title{Direct Analysis in Real Time Mass Spectrometry (DART-MS) Study of Amino Acids in the Vacuum Stable Liquids EMI-BF$_4$, EMI-Acetate, DEMA-OTf, DEMA-MsO, and Glycerol}

\author[1]{Zach Ulibarri}
\author[2]{Samantha E. Adamski}
\author[3]{Amy E. Hofmann}
\author[1]{Elaine M. Petro\thanks{Corresponding author: epetro@cornell.edu}}

\affil[1]{Sibley School of Mechanical and Aerospace Engineering, Cornell University, Ithaca, New York, USA}
\affil[2]{Department of Biology, Lycoming College, Williamsport, Pennsylvania, USA}
\affil[3]{NASA Jet Propulsion Laboratory, California Institute of Technology, Pasadena, California, USA}

\date{}

\begin{document}

\begin{abstract}

The detection of organics and potential biomolecules is an important goal of planetary science and astrobiology spacecraft missions, and many types of analytical chemistry for studying organics depend upon liquid solvents as working fluids. However, most liquids are difficult to maintain in the vacuum of space, complicating their use and increasing spacecraft mass and cost. Here we present a study using Direct Analysis in Real Time mass spectrometry (DART-MS) as a rapid screen for the detectability of histidine, glycine, arginine, lysine, and aspartic acid at mole fractions of about $5\times10^{-4}$ to $10^{-3}$ in five vacuum-stable liquids: EMI-BF$_4$, EMI-acetate, DEMA-MsO, DEMA-OTf, and glycerol. In all of the ionic liquids, the solvent cation dominated the spectra and no amino acid was detected, apart from a weak histidine signal in EMI-BF$_4$ at 350$^\circ$ C (median 5\% of the normalized peak intensity). The protic DEMA ionic liquids did not enhance detection, and their ionic components instead dominated the mass spectra. Aspartic acid was not detected in any liquid. In glycerol solutions, histidine, glycine, arginine, and lysine were all detected, and histidine was frequently a dominant line. We thus find glycerol to be the most promising of the liquids tested for organic detection with an ideal operating DART helium temperature of 350--400 $^\circ$C. We also found no clear evidence of water uptake or aqueous alteration in the ionic liquids, which was a secondary goal relevant to ionic liquid spacecraft electric propulsion.

\end{abstract}

\maketitle

\section{Introduction}

The search for extraterrestrial life and the characterization of organic inventories across the solar system depends on the ability of spacecraft to accurately detect and analyze organic compounds \cite{hendrix2019nasa,cable2020enceladus}. However, the environment of space poses many unique challenges for in-situ chemical analysis, one of which is the difficulty of maintaining the functionality and stability of liquid solvents under vacuum conditions. Traditional solvents, such as acetonitrile and methanol \cite{keppel2010use,kruve2017predicting}, used in numerous Earth-based analytical techniques such as electrospray ionization mass spectrometry (ESI-MS) and capillary electrophoresis (CE) are difficult to use in the vacuum of space: they require the spacecraft to carry along massive and bulky pressurized vessels, expendable solvent and pressurizing storage tanks, and associated valves and other hardware.

In laboratory settings, ESI is a preferred method of ionization for high-mass, fragile biomolecules because it is comparatively gentle and thus minimizes molecular fragmentation relative to other techniques \cite{fenn1989electrospray,fenn2003electrospray,verenchikov2023electrospray}. Despite the value of ESI-MS to the study and characterization of organic chemistry, no ESI-MS system has ever flown on a spacecraft, partly due to the size and cost requirements of operating with traditional volatile solvents \cite{arevalo2020mass}. Vacuum stable solvents thus may provide a unique opportunity to adapt analytical techniques critical to modern laboratory science for space exploration by enabling fluid operation in vacuum or very-low-pressure environments.

One plausible method for a spacecraft vacuum ESI-MS (vESI-MS) is to adapt existing vESI electric propulsion (EP) technology \cite{cogan2023electrospray}. These vESI EP systems use ionic liquids as propellants emitted at very high velocity and have the potential to enable exceptionally efficient spacecraft propulsion \cite{petro2020characterization}.

Ionic liquids are ionically-coupled anion-cation salts that exist in liquid form at ambient temperatures and pressures \cite{angell2012ionic}. Owing to their unique physical, thermal, and electric properties \cite{singh2020ionic}, ionic liquids are studied in a wide variety of contexts ranging from use as advanced lubricants in challenging or harsh environments \cite{bermudez2009ionic} to the carbon dioxide management in spacecraft life support systems \cite{tata2026supported}. Many have negligible vapor pressure, making them vacuum stable and thus attractive as vESI EP propellants. That same property makes them attractive as potential solvents for in situ organic analysis, where a liquid must survive in vacuum without bulky pressurized hardware. However, a useful working fluid for organic detection must also readily promote the ionization of any dissolved or suspended organic analytes. This necessitates the identification and characterization of vacuum-stable liquids that can effectively function as solvents for organic detection.

\begin{figure}[t]
    \centering
    \includegraphics[width=\columnwidth]{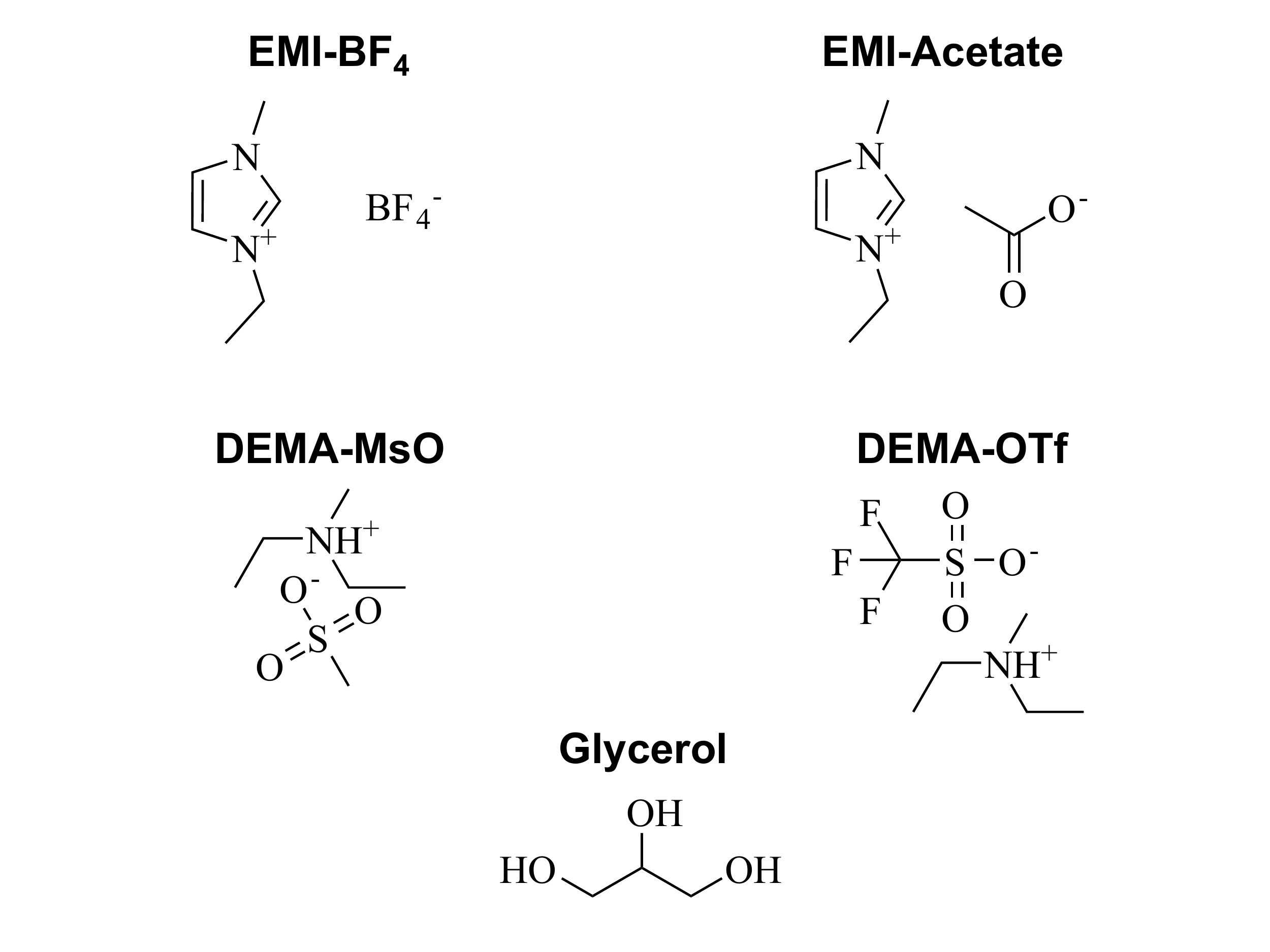}
    \caption{Skeletal structure formulae of the vacuum stable solvents used in this study.}
    \label{pubchem}
\end{figure}

Here we investigate the performance of several vacuum-stable liquids with a rapid screening method to qualitatively compare ionization efficiency and analyte detectability across a diverse set of vacuum stable solvents.  We scanned for amino acids in a selection of ionic liquids and glycerol, another vacuum stable solvent which has been previously used with  vacuum electrohydrodynamic ionization when doped with salts to increase conductivity \cite{simons1974electrohydrodynamic}. We performed measurements with a Direct Analysis in Real Time Mass Spectrometry (DART-MS) \cite{cody2005direct} at the Cornell University Chemistry Mass Spectrometry Facility. DART-MS works by placing samples into a stream of heated gas, typically helium or nitrogen, which is flowing into the vacuum inlet of a mass analyzer, in our case a Thermo Fisher Exactive Orbitrap. Because the measurement is performed with handheld samples in ambient conditions, it is a rapid means of studying many different samples with the high mass resolution Orbitrap. 

Skeletal structures of the selected vacuum stable liquids are shown in Fig. \ref{pubchem}. The preliminary characterization detailed in the present study is intended to highlight any vacuum stable solvents that are particularly promising not only for use in prototype vESI-MS systems, but potentially other future spacecraft instruments that may wish to make use of vacuum stable solvents for the detection of biosignature or other organics.

\citet{mazzotta2013direct} studied ionic liquids with DART-MS, finding intact cations to be the dominant mass spectral lines observed for various species but also that not all anions were readily observed in the spectra. \citet{gross2014high} also studied ionic liquids with DART-MS, looking in particular for high-mass clusters of the ionic constituents. Both studies examined the pure ionic liquids, and neither attempted to use DART-MS to find analyte organic species within the ionic liquids. To our knowledge, our present work is the first to use DART-MS to identify organic species dissolved in ionic liquids.

We further hypothesized that DART may be a quick way to qualitatively characterize contamination or aqueous alteration and degradation of ionic liquids. Water contamination is a persistent problem in EP laboratories \cite{conroy2009water}, and thus a quick means of assessing water content could be a useful laboratory diagnostic. Ionic liquids are also being developed as solvents for in situ processing of lunar regolith, where residual volatiles are removed by vacuum baking and water recovery from the liquid is a design goal \cite{rexhepi2023metal}, so a rapid vacuum-compatible water-sensitive compositional metric could therefore be useful in such studies. As a diagnostic for aqueous alteration for ionic liquid studies, we find a null result; DART-MS is highly sensitive to the ionic constituents of the liquid, but we found no strong mass spectral signs of hydrolysis or water retention.

For organic detection, we find that glycerol clearly stands out as a vacuum stable working fluid. Whereas the amino acids are uniformly not detected or detected only with weak, intermittent signatures in the ionic liquids, the amino acids in glycerol are robustly detected in the spectra at a wide variety of DART gas temperatures; in many of the spectra, the amino acids are by significant margin the highest intensity lines.

\section{Experimental Methods}

\subsection{Selection and Preparation of Solvents and Analytes}

\subsubsection{Vacuum Stable Liquids}

The ionic liquid 1-Ethyl-3-methylimidazolium tetrafluoroborate (CAS  143314-16-3, henceforth EMI-BF$_4$, also known as EMIM-BF$_4$ and C2mim-BF$_4$) is used as a propellant in vESI EP systems, where charged ionic droplets are electrostatically propelled at high velocity to maneuver small spacecraft, such as cubesats and nanosatellites \cite{ziemer2004microthrust,petro2020characterization}. EMI-BF$_4$ is an attractive species for such propulsion devices because it is electrically conductive, does not require an ionization energy cost for the spacecraft's power system (unlike other EP thrusters such as Hall effect thrusters), and has negligible vapor pressures which enable vacuum operation in liquid form. This liquid was studied primarily because of our expertise with it and has well-characterized firing characteristics in vESI EP operation.

1-Ethyl-3-methylimidazolium acetate (CAS 143314-17-4, henceforth EMI-Acetate) was selected because it is an ionic liquid that has the rare ability to directly dissolve amino acids and is related to EMI-BF$_4$.

Diethylmethylammonium methanesulfonate (henceforth DEMA-MsO, CAS 945715-44-6, also known as DEMA-OMs) and Diethylmethylammonium triflate (DEMA-OTf, CAS 945715-39-9) are protic ionic liquids, which can be used as liquid electrolytes for proton transfer in high temperature, anhydrous environments \cite{thimmappa2020diethyl,pan2020physicochemical,greaves2008protic}. DEMA-OTf in particular was used by \citet{fujiwara2015increasing} as the primary beam in secondary ion mass spectrometry (SIMS), where the protic structure was shown to promote the detection of amino acids in time-of-flight (TOF) MS by enabling greater rates of protonation of analytes relative to non-protic ionic liquids \cite{fujiwara2015increasing,fujiwara2017time}. DEMA-MsO is different from the other solvents studied here in that it was a mixture of both liquid and solid constituents at ambient conditions. Thus, the DEMA-MsO was stirred until a uniform texture was obtained. The DEMA-MsO was miscible with the water solutions, and after evacuation, the resulting mixture was visibly uniform, unlike the pure DEMA-MsO itself. This would imply that the hydrated DEMA-MsO is vacuum stable as a liquid even as our pure DEMA-MsO sample itself was quasi-stable as a liquid at room temperature.

Glycerol (CAS 56-81-5) was chosen because it is a vacuum stable liquid that is well-characterized, chemically much simpler than the ionic liquid solvents (which could reduce the likelihood of false positives in astrobiology missions, further discussed in Sec. \ref{sec:frag}), and it is sometimes used in CE, among other biologically-relevant instrumentation \cite{liang2001effect,cheng1994glycerol,mitchelson2001capillary}. Importantly, it has also been used in a vESI-like electrohydrodynamic ionization scheme by \citet{simons1974electrohydrodynamic}, indicating that it may be particularly attractive for the vESI-MS.

\subsubsection{Amino Acids}

Amino acids were chosen as the organic analytes because they are readily available, well-characterized, and, as the constituents of proteins, fundamentally required for terrestrial biology. The simplest amino acids (most notably glycine and alanine) are commonly found in abiotic inventories such as the Murchison meteorite \cite{kvenvolden1971nonprotein,engel1997isotopic} or returned asteroidal samples \cite{glavin2025abundant,naraoka2023soluble}. The more complex structures are thermodynamically disfavored in random production and thus less abundant in abiotic systems, with some never having been observed in abiotic inventories \cite{higgs2009thermodynamic}. Furthermore, most amino acids are composed of either right- or left-handed structures, which are created in equal proportion abiotically. Terrestrial life, however, only uses left-handed amino acids, creating a chirality disparity in biotic systems. Thus, in some contexts, the relative abundances of amino acids and/or chirality measurements may be used as biosignatures for life detection \cite{cable2020enceladus}.

The amino acid histidine was chosen because it is a comparatively complex amino acid that is disfavored in abiotic production \cite{higgs2009thermodynamic}, and it has been studied in the context of astrobiology, both by some of us \cite{ulibarri2023detection, schulze2022effect} and by other groups \cite{burke2023detection,klenner2024identify} in the context of hypervelocity impact ionization instruments \cite{cable2021plume}.

Glycine was chosen because it is the simplest amino acid, and it can be used as a baseline reading of amino acid content in biotic and abiotic systems. That is, the detection of more complex amino acids like histidine on extraterrestrial objects would be interesting, but not a clear biosignature. However, detection of significant quantities of the more complex amino acids relative the abundance of the simpler ones like glycine and alanine would be an important observation with significant astrobiological implications \cite{cable2020enceladus,davila2014chance,mckay2018enceladus}. As such, glycine was studied both by itself and in solution with histidine.

Lysine and arginine were chosen because, like histidine, they are complex enough to not be observed in abiotic inventories such as meteorite samples \cite{higgs2009thermodynamic}, they are comparatively easy to ionize, and they have also been studied in astrobiological contexts \cite{klenner2024identify}. For expediency, these two amino acids were studied simultaneously in a single solution. 

All the ionic liquids used in this study were acquired from Iolitec and were $>98\%$ purity, except for EMI-acetate, which was $>95\%$ purity, and reagent grade glycerol at $>99.5\%$ purity was sourced from Ricca Chemicals. The amino acids were sourced from Sigma Aldrich.

\subsubsection{Preparation}

We created $\sim$5 mM aqueous solutions of histidine, aspartic acid, glycine, and a dual mixture of lysine and arginine that we then mixed with the vacuum-stable liquids. We then evacuated the mixtures overnight under roughing pump in our EP vacuum chamber \cite{cogan2023electrospray}. Samples were prepared so as to create amino acid-to-vacuum-stable-liquid mole fractions of roughly $10^{-3}$ for the ionic liquids and $5\times10^{-4}$ for glycerol, as detailed in the Results, Table \ref{tab:detection}. During evacuation, much of the water from the original solutions evaporated away, leaving solutions of the amino acids in the vacuum-stable liquids. With the EMI-BF$_4$ mixtures, the amino acids could be seen as a white suspension of fine grained particles rather than a clear solution as observed with the other liquids.

\subsection{Experiment}
We attempted to follow the methodology of \citet{gross2014high}, wherein samples of the evacuated mixtures were applied to a metal mesh for use in the DART-MS. However, inconsistent results were achieved with this method, as the mass spectra changed dramatically with minute changes in mesh position or angle, so we adopted a simpler facility-standard method of dipping a borosilicate capillary tube into the mixtures which we then held into the DART extraction area.

To probe the differing optimal temperatures for each species, we studied each sample as a function of temperature. Thus, for each solution, a temperature sweep was taken across 50-550 $^\circ$ C in discrete 50 $^\circ$ C increments with fresh samples and capillaries using a helium working gas. At each temperature, a fresh sample was manually held into the inlet for $\sim$10 seconds, with 1.1 spectra saved per second. We then removed the sample, increased the DART temperature by 50 $^\circ$ C and waited for it to stabilize, and then repeated the process.

Mass spectra were acquired on a DART-SVP (Direct Analysis in Real Time- Standardized Voltage and Pressure) ion source (IonSense, Saugus, MA) coupled to an Exactive Orbitrap mass spectrometer (Thermo Scientific, Bremen, Germany). All sessions used a resolution setting of 100,000, an automatic gain control (AGC) target of $5 \times 10^{5}$, one microscan, no lock mass, and a capillary temperature of 250$^\circ$. Capillary, tube lens, and skimmer voltages were 30, 115, and 18 V respectively. 

Data was taken intermittently across several sessions spanning a few months. A detailed table of the various data taking sessions, their samples, and mass offsets is in included in the Results, Table \ref{tab:detection}.

\subsubsection{Software}

Spectra were recorded using the Thermo Fischer Exactive Tune software, with data saved through the Xcalibur data system as proprietary RAW files. These RAW files were converted to the open source mzML format using the ThermoRawFileParser package \cite{hulstaert2020thermorawfileparser} for downstream analysis with custom software.

Each run contains many scans taken while the sample was not in the inlet. We thus selected the scans rising above a 25\% baseline-to-maximum total ion current (TIC) from each run to find only higher intensity spectra when the sample was actually in the inlet region. This threshold was found empirically.

Because the DART is part of a user facility, the spectra contain many background lines from earlier samples and the atmosphere, especially at low temperatures. We then searched each selected scan for the expected ions: protonated amino acids and their sodium, potassium, ammonium, glycerol and ionic-liquid-cation adducts and protonated dimers, diagnostic fragments identified by \citet{zhang2019revisiting}, the ionic liquid cations, anions, dimers and trimers defined below, glycerol ions with their dehydration and ammonium products, known background lines, and water-related ions: protonated water clusters, hydrated cations, and clusters. The mass scale of the Exactive drifted by 0.2 to 1.3 mDa between different days, so we calculated a per-run offset from known ions and removed it before matching, detailed in Table \ref{tab:detection}.

We counted a species as detected when the intensity of the profile peak within 6 ppm of the expected mass exceeded ten times the local background in at least 25\% of the selected scans (to exclude single-scan spikes) and exceeded twenty times the local background in at least one scan (to ensure it was a strong, clear signal). The local background was the median of the nonzero profile intensities within $\pm$0.4 Da of the expected mass, excluding the matching window, with a floor of 50 counts. Mass lines were tagged as ``dominant'' if they had the highest intensity of at least 10\% of the selected scans. This 10\% threshold was arbitrarily chosen, but it should be noted that our results are largely insensitive to thresholds between 5 and 50\% as further detailed in Sec. \ref{sec:thresh}.

Ionic liquids are composed of anion--cation pairs and produce ion lines in various cluster combinations. Within the EP community these are referred to as monomers (the anion or cation, depending on MS polarity), dimers (the neutral anion--cation pair plus either the anion or the cation), trimers (two neutral pairs plus either ion), and higher-mass clusters \cite{romero2003source}. We searched explicitly for these clusters alongside the other expected products. Negative mode runs were acquired only for the DEMA liquids, where we searched for the deprotonated and chloride-adducted amino acids and for the solvent anions. Because the Exactive cannot study masses below 50 AMU, we did not consider any fragments below this mass.

We did not attempt to identify most background lines, and we label them ``Unknown'' in the plots. We label as ``Known background'' the lines we could attribute: dimethylsulfoxide (DMSO, CAS 67-68-5) at protonated mass 79.0212, N,N-dimethyl formamide (DMF, CAS 68-12-2) at mass 74.0600 (both commonly seen in the user facility), the solvent cations EMI$^+$ and DEMA$^+$ when they appear as contaminants in other solutions, and the lines at $m/z$~83.060, 100.076 and 110.081 that appear in control samples, called blanks. The blanks were created by mixing analyte-free deionized water with glycerol and evacuating it in the same manner as the samples with amino acids. Blanks were studied with glycerol only, since it is only with this liquid that the amino acids were detected. To decide which lines belong to an analyte, we compared each amino acid run with blanks and rejected peaks with large mass errors.

Finally, we combined the selected scans of one sample for each solvent--analyte combination into a single 3D plot. For every selected scan we picked the peaks above 3\% of the highest intensity peak and plotted each at the scan temperature. We overlaid all scans with partial transparency, which highlights the lines that are frequent and abundant and lets us characterize the whole parameter space of each sample at a glance.

\section{Results}
\begin{table*}[htbp]
\caption{\textbf{Samples and detection of protonated amino acids.} ``Mole fraction'' is amino-acid-to-liquid, computed from the weighed masses and uncertain by about 10\%. Temperatures are DART gas temperatures ($^\circ$ C). ``Detected'' lists temperatures at which the $[\mathrm{M{+}H}]^+$ line was found in positive mode, and ``dominant'' those at which it was the highest-intensity peak in at least 10\% of the selected scans. $^\dagger$ marks detections that never exceeded 10\% of the highest-intensity peak. $^\S$ marks samples for which full temperature sweeps were run but only the listed temperatures were saved. No line was dominant in any scan during these sweeps, as judged during acquisition, but detection at the unsaved temperatures cannot be excluded. ``Offset'' is the median per-run mass offset removed before matching (mDa). No amino acid was detected in negative mode.}
\label{tab:detection}
\fontsize{7}{8.4}\selectfont
\setlength{\tabcolsep}{4pt}
\renewcommand{\arraystretch}{1.1}
\begin{tabular}{@{}>{\raggedright\arraybackslash}p{0.075\textwidth}>{\raggedright\arraybackslash}p{0.125\textwidth}>{\raggedright\arraybackslash}p{0.045\textwidth}>{\raggedright\arraybackslash}p{0.060\textwidth}>{\raggedright\arraybackslash}p{0.115\textwidth}>{\raggedright\arraybackslash}p{0.155\textwidth}>{\raggedright\arraybackslash}p{0.115\textwidth}>{\raggedright\arraybackslash}p{0.095\textwidth}>{\raggedright\arraybackslash}p{0.045\textwidth}@{}}
\toprule
 & & & & \multicolumn{3}{c}{Positive mode} & Negative & \\
\cmidrule(lr){5-7}
Liquid & Amino acid & Date & Mole fraction ($10^{-4}$) & Temperatures & Detected & Dominant & Temperatures & Offset \\
\midrule
EMI-BF$_4$ & Arginine (with Lys) & 16 Aug & 11 & 200, 450 & none$^{\S}$ & none & -- & 0.38 \\
 & Lysine (with Arg) & 16 Aug & 11 & 200, 450 & none$^{\S}$ & none & -- & \\
 & Aspartic acid & 16 Aug & 11 & 450 & none$^{\S}$ & none & -- & 0.24 \\
 & Histidine & 11 Oct & 11 & 50--550 & 300$^{\dagger}$, 350 & none & -- & 0.77 \\
\midrule
EMI-Acetate & Histidine & 10 Oct & 10 & 50--550 & none & none & -- & 0.54 \\
\midrule
DEMA-MsO & Histidine & 30 Sep & 9.3 & 50--550 & none & none & 50--550 & 1.25 \\
 & Aspartic acid & 30 Sep & 10 & 50--550 & none & none & 50--550 & 1.22 \\
 & Histidine & 8 Oct & 9.8 & 50--200, 300--500 & none & none & 50--500 & 0.40 \\
\midrule
DEMA-OTf & Arginine (with Lys) & 27 Sep & 13 & 50--550 & none & none & 100--550 & 1.09 \\
 & Lysine (with Arg) & 27 Sep & 13 & 50--550 & none & none & 100--550 & \\
 & Aspartic acid & 27 Sep & 13 & 100--500 & none & none & -- & 1.09 \\
 & Histidine & 8 Oct & 13 & 100, 200--500 & none & none & -- & 0.46 \\
\midrule
Glycerol & Arginine (with Lys) & 18 Sep & 5.4 & 50--250, 350--450 & 350, 450 & none & -- & 0.93 \\
 & Lysine (with Arg) & 18 Sep & 5.6 & 50--250, 350--450 & 250, 350--450 & 250, 350--400 & -- & \\
 & Aspartic acid & 18 Sep & 5.1 & 50, 150--500 & none & none & 300, 400, 500 & 0.92 \\
 & Histidine (with NaCl) & 19 Sep & 4.8 & 50--500 & 300--500 & 350--500 & -- & 0.48 \\
 & Histidine & 25 Sep & 5.1 & 50--500 & 300--500 & 350--500 & -- & 0.90 \\
 & Histidine & 10 Oct & 5.0 & 50--550 & 250--300$^{\dagger}$, 350--550 & 400--550 & -- & 0.68 \\
 & Glycine & 17 Dec & 4.9 & 50--550 & 300--550$^{\dagger}$ & none & -- & 0.60 \\
 & Blank (glycerol and water) & 17 Dec & -- & 50--550 & -- & -- & -- & 0.59 \\
 & Glycine & 20 Dec & 5.0 & 50--500 & 150--200$^{\dagger}$, 250--500 & 300--450 & -- & 1.06 \\
 & Histidine (with Gly) & 20 Dec & 5.0 & 50--500 & 250$^{\dagger}$, 300--500 & 300--500 & -- & 1.02 \\
 & Glycine (with His) & 20 Dec & 4.9 & 50--500 & 150$^{\dagger}$, 200--500 & 350 & -- & \\
 & Blank (glycerol and water) & 20 Dec & -- & 50--500 & -- & -- & -- & 1.08 \\
\bottomrule
\end{tabular}
\end{table*}

\label{results}
\begin{figure*}
\centering
\includegraphics[width=\textwidth]{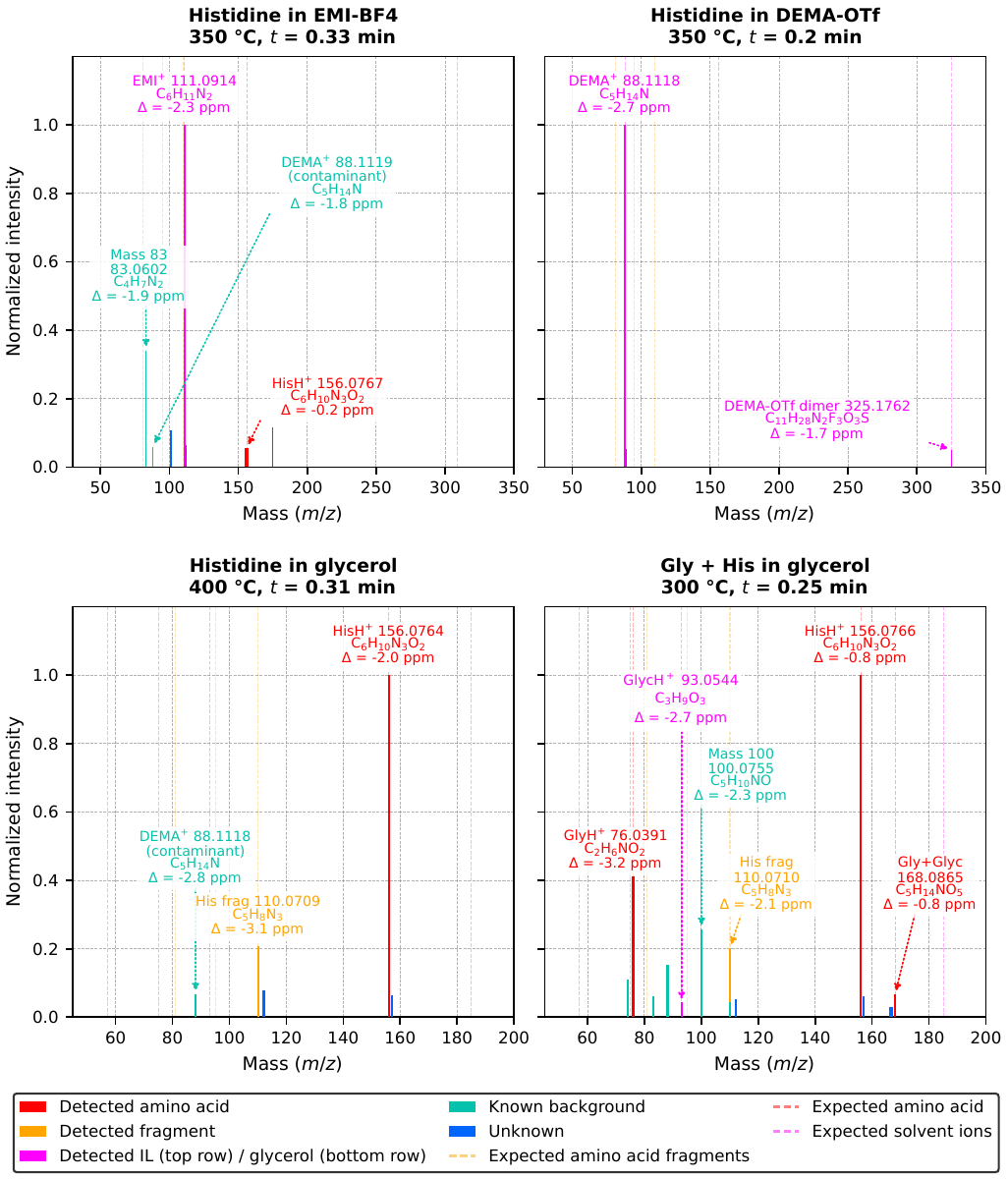}
\caption{Representative DART-MS spectra of histidine in (top left) EMI-BF$_4$, (top right) DEMA-OTf, and (bottom left) glycerol, and of glycine and histidine together in glycerol (bottom right), normalized to highest intensity peaks. Dashed vertical lines mark expected masses, and $\Delta$ represent mass error.}
\label{fig:spectra}
\end{figure*}

The detections as a function of temperatures for each solvent--analyte pair are summarized in Table \ref{tab:detection}. Representative spectra are shown in Fig. \ref{fig:spectra}, where each panel shows one analyzed scan normalized to its highest intensity peak. In EMI-BF$_4$ at 350$^\circ$C, the solvent cation was the highest intensity peak in 9 of 12 analyzed scans. The scan shown in Fig. \ref{fig:spectra} has the histidine line at 5.6\% of the normalized peak, close to the median of 4.8\% over the run. In DEMA-OTf at 350$^\circ$C, the cation was the highest intensity peak in all 10 analyzed scans and none contained a histidine line. In glycerol at 400$^\circ$C, histidine was the highest intensity in 9 of 20 scans (median 76\% of the normalized peak). The glycine--histidine mixture is shown at 300$^\circ$C, the temperature at which histidine first becomes dominant in that mixture. In that run, histidine was the dominant peak in 10 of 17 scans. The shown spectrum has glycine at 41\% of the normalized peak, the run median.

\begin{figure*}
\centering
\includegraphics[width=\textwidth]{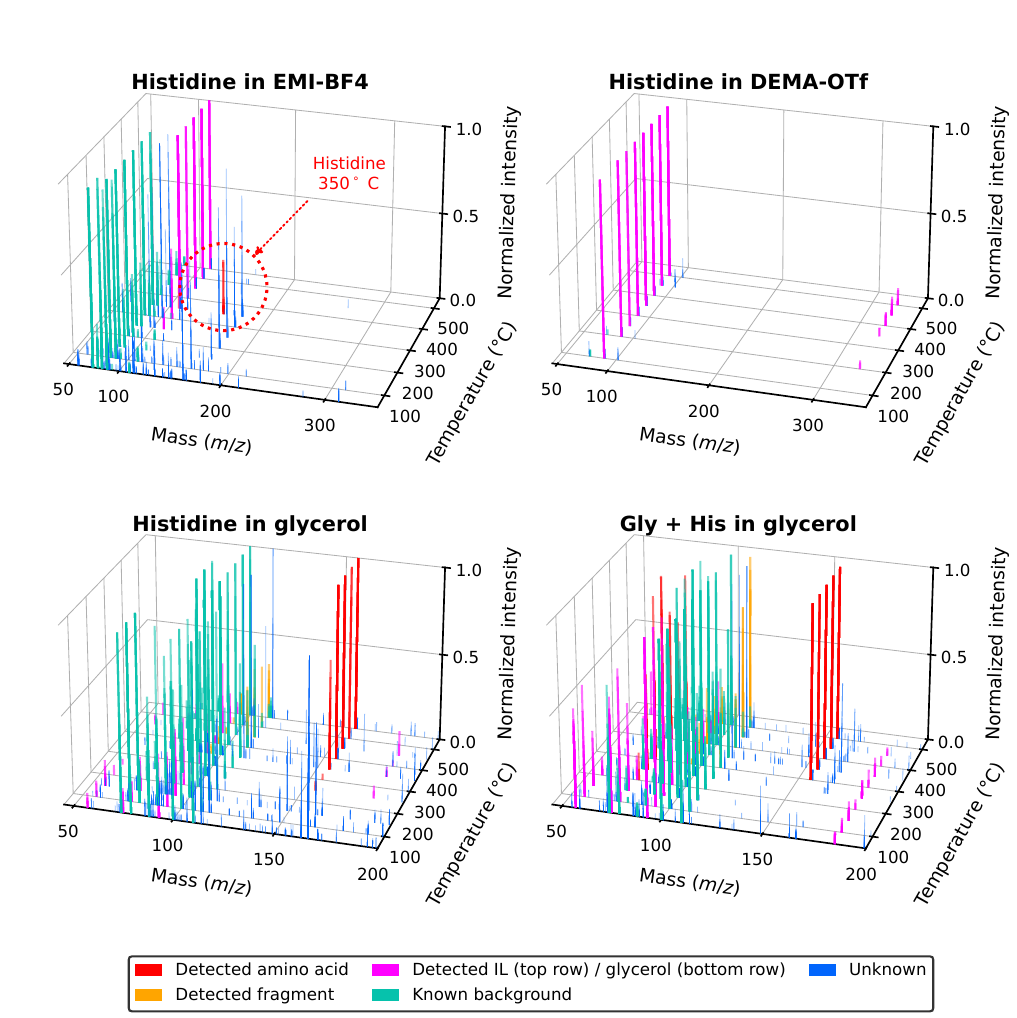}
\caption{Temperature sweeps of histidine in EMI-BF$_4$ and DEMA-OTf (top), and of histidine and glycine plus histidine in glycerol (bottom). Each line is one mass line in one analyzed scan, plotted at its DART gas temperature and normalized to the highest intensity line in that scan. The red dotted circle marks the single histidine detection in EMI-BF$_4$.}
\label{fig:temps3d}
\end{figure*}

Examples of the temperature sweeps are shown in Fig. \ref{fig:temps3d} with a red circle highlighting the lone amino acid detections in an ionic liquid, at 350$^\circ$C in EMI-BF$_4$. 

No amino acid appeared in DEMA-MsO, DEMA-OTf, or EMI-Acetate at any temperature, nor in any negative mode run. In EMI-BF$_4$, histidine appeared at 350$^\circ$ C in 92\% of the selected scans, but with a median of only 5\% of the normalized peak and a maximum of 32\% in two adjacent scans, and with a trace at 300$^\circ$ C (2\%). Arginine, lysine, and aspartic acid in EMI-BF$_4$ were not detected at the temperatures for which full analysis exists (200 and 450$^\circ$ C, or 450$^\circ$ C). Full temperature sweeps were run for these samples but the other temperatures were not saved. We are confident that no amino acid line was dominant in these missing scans during the sweeps, but detection at these unsaved temperatures cannot be excluded. In every ionic liquid, the solvent cation was the highest intensity peak in most scans (79 to 100\% for the DEMA liquids and EMI-Acetate, 44 to 92\% for EMI-BF$_4$).

Water-related ions were searched for in all saved ionic liquid runs (91 positive- and 45 negative-mode runs). No protonated water clusters, anion hydrates, or hydrolysis products of \ce{BF4-} (\ce{BF3OH-}, \ce{BF2(OH)2-}) appeared, and the hydrated cation--anion clusters were absent, with the hydrated EMI-BF$_4$ dimer reaching at most 0.5\% of the normalized peak. The only candidate was a weak line matching $[\mathrm{EMI{+}H_2O}]^+$ at $m/z$ 129.102, which met the detection criterion in 2 of 11 EMI-Acetate runs (100 and 150$^\circ$ C, up to 12.5\% of the normalized peak) and in 2 of 14 EMI-BF$_4$ runs (450$^\circ$ C, up to 8\%).

In glycerol, histidine appeared in all four samples from 300$^\circ$ C, with a trace at 250$^\circ$ C in two of them. It was a dominant line from 350$^\circ$ C upward when in solution by itself, with or without NaCl, and from 300$^\circ$ C upward in the glycine--histidine mixture.

Glycine appeared in both its glycerol samples, being detected at 150$^\circ$ C and above in one sample, but only at 300$^\circ$ C and above in the other. It is unclear why such a disparity was observed when the samples had such similar mole fractions. It was a dominant line from 300 to 450$^\circ$ C when in solution by itself, and at 350$^\circ$ glycine--histidine mixture. In glycine-containing runs at 300$^\circ$ C and above, a line at $m/z$ 168.087, matching the glycine--glycerol adduct appeared in 12 of 16 runs and in none of 13 blank runs.

Arginine appeared at 350 and 450$^\circ$ C, not being detected in between at 400$^\circ$ C, and lysine at 250 and 350 to 450$^\circ$ C (dominant at 250 and 350 to 400$^\circ$ C). Aspartic acid never appeared in any solvent in either polarity. 

The histidine fragment at $m/z$ 110.071 (\ce{C5H8N3+}) appeared from 300$^\circ$ C in three samples and from 350$^\circ$ C in the fourth, always at or after the first clear detection of the protonated molecule. It was the highest intensity peak in two samples and 41 and 76\% of the normalized intensity in the other two. A line at $m/z$ 112.087 (\ce{C5H10N3+}) appeared in all 23 histidine-containing glycerol runs at 300$^\circ$ C and above and in none of 13 blanks or 11 glycine-only runs. It also appears in the 350$^\circ$ C arginine data, consistent with being reported as a fragment of arginine in \citet{zhang2019revisiting}. Known histidine fragments at $m/z$ 95.060 (\ce{[C5H7N2]+]}) and 81.045 (\ce{[C4H5N2]+} were also present in the histidine data and absent from blanks. 

With glycerol, there is a $\le$5 second period before amino acids become strong mass lines. Histidine and its fragments grow over the first 2 to 5 seconds while a line at $m/z$ 100.076 declines relative to them, so early spectra can show lines that later fade. 

The line at $m/z$ 100.076 is a prominent feature of the glycerol spectra. It was the highest intensity peak in 85\% of the selected scans of the glycerol/DI-water blanks at 300$^\circ$ C and above, appeared in 93\% of glycerol runs at 300$^\circ$ C and above, and was essentially absent from the ionic liquids. In the blank, its absolute intensity was about five times that in the glycine--histidine runs measured on the same day and 1.1 to 1.7 times that in the glycine-only runs.

\section{Discussion}
\label{discussion}

\subsection{Glycerol versus the Ionic Liquids}
Amino acids appeared readily in glycerol solution mass spectra and almost never in the ionic liquids (Table \ref{tab:detection}), meaning that glycerol clearly stands above the other liquids for the purposes of amino acid detection in the present study. The ionic liquids are by definition composed of ionic constituents, and the solvent cation was the dominant line in most scans, whereas glycerol is a highly stable neutral molecule. A simple explanation is thus that in charge competition, the ionic liquids easily overpower the spectra and drown out any other lines, while for glycerol the analytes are more likely to successfully protonate. Such an explanation is strengthened by the fact that the protic DEMA liquids, which promoted protonation in secondary ionization in \citet{fujiwara2015increasing}, did not induce easier detection in the present study.

Rather, the DEMA species almost completely dominated the spectra, drowning out detection of even most background masses and becoming a significant contaminant in later data taking sessions, some days or even weeks later. Thus, the DEMA ionic constituents were so easy to ionize, they made it nearly impossible to see anything else. Because small lines are rejected based on the background surrounding them, not the normalized intensity, this result is not an artifact of our downselection process. Even in the DEMA spectra, over a hundred other lines are observed, but none that match the amino acid parent, fragments, or relevant clusters thereof.

The one ionic liquid detection, histidine in EMI-BF$_4$ at 350$^\circ$ C, is weakly detected. It is present in 11 out of the 12 analyzed spectra at that temperature. Its median normalized intensity is 5\%, but in two adjacent spectra, it is 32\% of the normalized intensity. It can be confirmed as a real line because the characteristic histidine fragment at 110.071 tracks it in its two high intensity spectra: the fragment-to-parent intensity ratios in these two spectra are 30-40\%, roughly the same as those in the glycerol data where histidine is observed very brightly.

It is interesting, however, that the histidine signal is repeatably measured in this temperature but then not observed at all even in adjacent temperatures (excepting a trace detection at $300^\circ$ C). It is important in this context to remember that with EMI-BF$_4$, the analytes were observed as a white, grainy suspension rather than the solutions observed with other liquids. It could thus be explained that the transient, inconsistent detection in this manner is the result of a burst of ions produced when a suspended grain reached some condition in which it produced significant ions, but otherwise this was difficult to achieve over the constant stream of EMI$^+$ cations.

It is also important to remember that the ionic liquid results are somewhat limited, since each pair has one or two samples and the data saved and fully analyzed for arginine, lysine and aspartic acid in EMI-BF$_4$ cover only a few temperatures, so a weak detection there cannot be excluded (Table \ref{tab:detection}). Nonetheless, these other temperatures were not saved during testing precisely because no amino acid signal could be observed during real-time analysis, so we can at minimum rule out their being easy to detect with this method.

\subsection{Detection in Glycerol}

In glycerol, the ideal detection temperature differs between amino acids. Histidine was a dominant line in the mass spectra from 350$^\circ$ C upward, both by itself and with NaCl, and from 300$^\circ$ C in the glycine mixture. It was dominant in a larger fraction of the spectra at 400$^\circ$ C and above, indicating slightly preferential ionization at the higher temperatures. Glycine was dominant at 300 to 450$^\circ$ C when by itself and at 350$^\circ$ C in the mixture. Lysine was dominant at 250 and 350 to 400$^\circ$ C. These results thus suggest an overall optimum temperature for detection of dissolved amino acids in glycerol at 350-400$^\circ$ C with helium as a working gas in DART-MS. It is interesting to note that \citet{mazzotta2013direct} and \citet{gross2014high} both found optimum results at temperatures of 400-500$^\circ$ C for entirely different objectives.

Histidine was detected in four independent samples taken over months of varied testing, with two in solution by itself, one with $\sim$ 1 g/L \ce{NaCl}, and one with glycine. When mixed with glycine, the results are largely consistent, although interestingly it is only in this solution that histidine was dominant at 300$^\circ$ C.

In the glycine--histidine mixture, glycine was dominant only at $350^\circ$ C, whereas it was in the glycine-only sample run on the same day in the 300-450$^\circ$ range, consistent with preferential protonation of the slightly more basic histidine.

It is particularly interesting that the histidine-in-glycerol solution with \ce{NaCl} showed the same detection temperatures as the salt-free samples, with no sodiated ions observed either for parent amino acid molecules or for fragment products. As sodium is a major contaminant in many positive-ion MS systems that complicates analysis in astrobiological contexts \cite{napoleoni2023mass,bryson2023development,seaton2023robust}, including our own \cite{ulibarri2026hypervelocity}, this may hint at an interesting avenue for future spacecraft instrumentation studies.

Aspartic acid was the only amino acid that was not detected at all, and it was the only amino acid added as the free acid (as opposed to an \ce{HCl} salt), which may indicate that detection requires a pre-protonated or salt form, although this was not explicitly tested.

\subsection{Background Lines and Fragments}
\label{sec:frag}

The line at $m/z$ 100.076 is a glycerol-associated background species and not an analyte product. It is the highest intensity peak in glycerol blanks, it is essentially absent from the ionic liquids. The time-dependent nature of 100.076 masks the early appearance of amino acid lines, but later fades and reveals them. Either the sample as a whole must reach a high temperature before histidine ionizes, or this competing glycerol product must first locally deplete. Compared to blank glycerol runs, the 100.076 signal is about five times weaker in the glycine--histidine runs from the same day but only 1.1 to 1.7 times weaker in the glycine-only runs. This observation suggests that histidine suppresses the signal, subject to the limitations of the limited sampling of the present study. Its normalized intensity stays near unity in the blanks, where nothing competes with it, so its fade in the sample runs reflects the growth of the amino acid lines. We tentatively assign it to protonated N-methyl 2-pyrrolidone (\ce{C5H10NO+}), a common solvent in cleaning products and plastics and a known MS contaminant species \cite{keller2008interferences}. The same formula would arise from loss of neutral \ce{CN2O} from histidine, which the literature does not report as a histidine fragment \cite{nisthist,zhou2017ionic,ulibarri2023detection}, and its presence in blanks rules out a histidine origin.

Some of us have studied histidine fragmentation in hypervelocity impact experiments \cite{ulibarri2023detection}, where histidine degraded to fragments at masses 81 to 83 and 110, with 110 identified as \ce{[M+H-H2O-CO]+} (\ce{C5H8N3+}, 110.07114 AMU) \cite{zhang2019revisiting}. A significant question in that study was if the 110 line was present as a pre-ionization fragment product. In the present glycerol data, the 110.071 fragment appears at or after the temperature where the protonated molecule is first clearly detected. This observation would suggest fragmentation during desorption and ionization rather than existence as a fragment in the aqueous solution. 

Other lines observed at $m/z$ 83.060 and 110.081 also occur in glycerol blanks. Mass 83.060 would match the histidine fragment \ce{[C4H7N2]+}, but also the EMI cation minus ethylene.  The 110.081 line has a mass error of 90 ppm from the expected 110.071, well outside of plausible error. It matches nicely with glycerol adducted with an ammonium ion (\ce{[C3H8O3 + NH4]+}, mass error -0.5 ppm) and appears in 97\% of glycerol blanks. Being well below the 50 AMU floor of the mass analyzer, it is unsurprising that \ce{NH4+} is not observed as corroboration. 

The signal at $m/z$ 83 is difficult to assign because the same species is a fragment of both histidine \cite{zhang2019revisiting,ulibarri2023detection} and EMI-based ionic liquids, where it is frequently seen in collisional fragmentation spectra \cite{hofheins2026secondary,bell2026experimental} and modeling results \cite{tahsin2025reactive}. Mass 95 (\ce{[C5H7N2]+]}) is similarly a fragmentation product of both histidine \cite{zhang2019revisiting} and EMI-based ionic liquids \cite{hofheins2026secondary,bell2026experimental}. That both 83 and 95 create overlapping fragment products that cannot be directly attributed to either histidine or EMI-based products further illustrates the difficulty of using such ionic liquids for biosignature analysis: the complex chemical nature of the liquid may create false-positive signals or complicate proper identification of relevant biosignatures. In this case, the common imidazole ring creates identical fragments. Thus, entirely aside from the present result of increased detectability of amino acid parent molecules, the relative simplicity of the glycerol molecular structure relative to the ionic liquids makes false-postives less likely for astrobiology spacecraft missions.

In a recent study where two samples of aluminum dust, one dosed with arginine and the other with histidine, were studied with hypervelocity impact ionization mass spectrometry \cite{ulibarri2026hypervelocity}, mass 112 was observed for both species in a high velocity subset of the data. It is not reported as a fragment of histidine in either \citet{zhang2019revisiting} or the NIST WebBook, but it is reported for arginine. In the present study, m/z 112.087 (\ce{C5H10N3+}) appeared in every glycerol sample containing histidine, sometimes as the highest-intensity peak, but in no blank, glycine or aspartic acid sample. It also appeared with histidine in DEMA-MsO at 500--550$^\circ$ C and in the Arg--Lys glycerol sample at 350 and 450$^\circ$ C. Its mass matches protonated histamine, the decarboxylation product of histidine.

\subsection{``Dominant'' Line Threshold}
\label{sec:thresh}

The choice of a 10\% threshold for the labeling of mass lines as ``dominant'' is arbitrary, but the results presented here are largely indifferent to this selection. Raising it from 5 to 25\% shifts the ends of the dominant temperature ranges in glycerol by at most one 50$^\circ$ C step. At 50\% the ranges narrow further and lysine is no longer dominant. The primary result of the present study is that the amino acids are significantly easier to detect in glycerol than in the other liquids. Because the amino acids were frequently the highest intensity peaks in the glycerol data, and they were nearly absent from and never the highest intensity peaks in the ionic liquid spectra, the arbitrary threshold selection is irrelevant to this primary result; it is merely useful as a simple, reproducible criterion to determine when the amino acids are the highest intensity lines in a temperature scan.

\subsection{Limitations and Transferability}
DEMA$^+$ is exceptionally easy to ionize and is clearly a contaminant in this study of our own origin. Carry-over also appears in negative mode, where triflate shows up in DEMA-MsO runs after the DEMA-OTf runs. EMI$^+$ also appeared sporadically in glycerol runs, reaching 39\% of the normalized peak in some spectra. Solutions were measured days apart, yet still often showed ionic liquid contaminant, especially with DEMA-based species. As such, full background characterization and the more important false-positive analysis was deemed impractical.

That glycerol stands above the ionic liquids for organic detection with DART is clear, but the transferability of this result towards the vESI-MS is not straightforward. The differing ionization mechanisms between the hot gas phase helium in DART and the electric field desorption of a vESI-MS mean that the results may not directly carry over to such an instrument. Nonetheless, the results, combined with previous studies of glycerol in electrohydrodynamic ionization \cite{simons1974electrohydrodynamic,ku2003electrohydrodynamic} suggest it may have some advantages over other vacuum-stable species, be it for the vESI-MS or other future spacecraft instrumentation, (e.g., \citet{creamer2017enhanced,seaton2023robust}).

\subsection{Water Uptake Diagnostic for Ionic Liquids}
A secondary aim was to test DART-MS as a rapid diagnostic for water uptake by ionic liquids. Little MS evidence of aqueous alteration was present in the data, although many of the simplest water products would be below the Exctive's $m/z$ floor of 50, and negative ion data was not taken for most species. The only candidate observation was a weak \ce{[EMI + H2O]+} line, seen at 100 and 150$^\circ$ C in EMI-Acetate, a strongly hygroscopic liquid, and at 450$^\circ$ C in two EMI-BF$_4$ runs. Our present results thus suggest that DART is not a good candidate for this diagnostic.

\section{Conclusion}

We used DART-MS as a rapid screen for the detectability of five amino acids in four vacuum-stable ionic liquids (EMI-BF$_4$, EMI-acetate, DEMA-MsO, and DEMA-OTf) and in glycerol, over DART helium temperatures of 50--550$^\circ$ C, although not all temperatures were fully studied for all species. Glycerol was the most favorable solvent for organic detection. Histidine, glycine, arginine, and lysine were all detected in glycerol. Histidine, glycine, and lysine were the highest intensity mass lines in some glycerol spectra at or above 300$^\circ$ C. 

In the ionic liquids, the solvent cation dominated the spectra, and the only amino acid detection was a weak histidine signal in EMI-BF$_4$ at 350$^\circ$ C and at trace level at 300$^\circ$ C. The protic DEMA liquids, which promote amino acid protonation in SIMS \citep{fujiwara2015increasing}, did not improve detection in DART, and their cations dominated the signal in nearly all of our relevant spectra. We tentatively attribute the contrast to competition for charge between the pre-formed solvent ions and the analytes, a competition that a neutral solvent such as glycerol avoids. The optimal DART temperature for the detection of amino acids in glycerol appears to be 350--400$^\circ$ C. Aspartic acid was not detected in any liquid, and it interestingly was the only analyte added as a free acid rather than as an HCl. In glycerol, the histidine fragment at $m/z$ 110.071 appeared at or after the first clear detection of the protonated molecule, which suggests that it forms during desorption and ionization rather than being present in solution, which is relevant to ongoing studies of histidine in hypervelocity impact ionization studies \cite{ulibarri2026hypervelocity}. Added NaCl had no visible effect on histidine detection, in contrast to many positive ion mass spectrometry studies \cite{napoleoni2023mass}.

As a rapid diagnostic for water uptake or hydrolysis in ionic liquids, DART-MS gave a null result. We found no clear evidence of aqueous alteration, only a weak $[\mathrm{EMI{+}H_2O}]^+$ line in a limited number of scans.

These conclusions carry important caveats. Most solvent--analyte pairs were represented by only one or two samples, each with a single capillary dip per condition, so a non-detection means only that no line rose above the background. Ionic liquid carry-over in the shared user facility prevented a full background and false-positive characterization. In addition, hot-gas ionization in DART differs from other mechanisms of ionization, particularly field-driven ionization in a vESI source, so these results may not transfer directly to spacecraft instrumentation. Nonetheless, the clear advantage of glycerol in the present study suggests it may have special utility as a working fluid in future spacecraft-based organic detection instruments.

\section{Acknowledgments}
The authors would like to thank Dr. Rob Hodyss of the NASA Jet Propulsion Laboratory for his assistance in suggesting N-methyl 2-pyrrolidone as an identification for mass line 100.076.

This work was supported in part by the Heising-Simons Foundation 51 Pegasi B Fellowship grant number 2024-5175. 

A portion of the work was supported by a grant from the Jet Propulsion Laboratory, California Institute of Technology, under contract with the National Aeronautics and Space Administration (80NM0018D0004).

This work made use of the Cornell University NMR (RRID:SCR\_028078) and Chemistry MS (RRID:SCR\_028079) Facilities.

\bibliography{zbibliography}

\end{document}